\documentclass[12pt,preprint]{aastex}

\shorttitle{Distance to PWN G54.1+0.3 and Search for its SNR Shell}

\shortauthors{Leahy, Tian and Wang}

\begin{document}

\title{Distance Determination to the Crab-like PWN G54.1+0.3 and Search for its Supernova Remnant Shell}

\author{Denis A. Leahy\altaffilmark{1}, Wenwu Tian\altaffilmark{2,1}, 
Q.D. Wang\altaffilmark{3} }
\altaffiltext{1}{Department of Physics \& Astronomy, University of Calgary, Calgary, Alberta T2N 1N4, Canada, leahy@iras.ucalgary.ca}
\altaffiltext{2}{National Astronomical Observatories, CAS, Beijing 100012, China, tww@iras.ucalgary.ca}
\altaffiltext{3}{Department of Astronomy, University of Massachusetts, 710 North Pleasant Street, Amherst, MA 01003, USA} 

\begin{abstract}
We discover a large-scale shell G53.9+0.2 around the Crab-like pulsar wind nebula
(PWN) G54.1+0.3 with 1420 MHz continuum VLA observations. This is confirmed by a new
 infrared image at 8 $\mu$m from the GLIMPSE Legacy Project, which reveals an intriguing 
infrared shell just surrounding the large radio shell. 
 We analyze the 21 cm HI absorption spectra and $^{13}$CO emission spectra towards PWN 
G54.1+0.3 and bright sources on both radio and IR shells. Continuous HI absorption up to 
the tangent point and absence of negative HI absorption features imply that PWN G54.1+0.3 has
a distance beyond the tangent point but within the Solar circle, 
i.e. 4.5 to 9 kpc. G54.1+0.3 is likely at distance of $\simeq$6.2 kpc due to 
the morphological association between the PWN and a CO molecular cloud at velocity of 
$\simeq$53 km/s, as revealed by high-resolution $^{13}$CO images. Based on the HI absorption
 spectrum and recombination line velocity ($\simeq$40 km/s) of the bright HII region 
G54.09-0.06, which is on the IR shell, the IR shell is likely located at a distance of 
$\simeq$7.3 kpc, which is also the distance of the associated large-scale radio shell. At this
 distance, the radio shell has a radius of $\sim$30 pc. 
The radio shell  may be thermal and lack IR emission due to dust destruction, 
or it may be nonthermal and
part of a newly found old SNR. In either case it is located at a different distance than 
PWN G54.1+0.3. 
 
\end{abstract}


\section{Introduction}
PWN G54.1+0.3 is a Crab Pulsar Wind Nebula (PWN) analog, with a young central pulsar PSR J1930+1852
spinning down to energize its PWN (Velusamy \& Becker, 1988). 
The Crab Nebula, so far, despite intensive searches (Frail et al. 1995)
shows no evidence for an outer shock (i.e. supernova remnant (SNR) shell)
from the supernova explosion which produced the pulsar. No satisfactory explanation
has been proposed for this lack of an outer SNR shell. Like the Crab, G54.1+0.3
also, as yet, has no identified outer SNR shell. Here we search for such a shell for
G54.1+0.3. We find an SNR-like shell around G54.1+0.3 and then study its properties to
determine if it is associated with G54.1+0.3.

The distance of PWN G54.1+0.3 is important for determination of the luminosity and 
energetics of the pulsar and its PWN and for comparison with other PWN.
The distance has been estimated by several times previously. 
Based on Chandra X-ray observations, Lu et al. (2002) estimated an absorption column density 
of about half the total Galactic absorption in the direction to G54.1+0.3, and therefore 
suggested a distance of $\sim$5 kpc. From an HI  absorption spectrum,
Weisberg et al. (2008) found PSR J1930+1852, which is the central pulsar of PWN 
G54.1+0.3, is at a distance of $>$5 kpc. Using an improved free-electron density/distance model
 (Cordes \& Lazio 2002), Camilo et al. (2002) gave a DM distance of $\le$8 kpc to PSR J1930+1852. %Assuming the SNR to be associated with the star-forming region G53.9+0.3, Velusamy \& Becker (1988) obtained the distance of $\sim$ 3.2 kpc. 
The distance uncertainty on the system is thus still large, and improved distance
 determination to the PWN system is needed. Recently, we have revised distances to several 
SNR/PWN or SNR/HESS source or SNR/CO cloud interaction systems with good precision by analyzing HI+CO
 spectra of these system (Leahy \& Tian 2008; Tian, Leahy \& Wang 2007; Tian et al. 2008). Here, we use these 
effective methods to constrain the distance to the PWN G54.1+0.3/ PSR J1930+1852 system, 
and to revise its X-ray luminosity. 

\section{Radio and Infrared Continuum Emission}
The radio continuum at 1420 MHz and HI-line emission data sets come from the Very Large Array
 (VLA) Galactic Plane Survey (VGPS), described in detail by Stil et al. (2006). The spatial
 resolution of the continuum images of SNR G54.1+0.3 is 1$^{\prime}$ (FWHM) at 1420 MHz. The 
synthesized beam for the HI line images is 1$^{\prime}$, the radial velocity resolution is 1.56 
km$/$s, and the rms noise is 2 K per channel. The short-spacing information for the HI spectral 
line images in the VGPS is from additional observations with the 100 m Green Bank Telescope of the NRAO.  
The CO-line ($J=1-0$) data set is from the Galactic ring survey (Jackson et al. 2006) made with 
the Five College Radio Astronomy Observatory 14 m telescope. The CO-line data in this paper
 have velocity coverage of -5 to 135 km/s, an angular sampling of 22$^{\prime\prime}$, radial 
velocity resolution of 0.21 km/s, and rms noise of $\sim$0.13 K.  

Figure 1 (left panel) gives the 1420 MHz continuum image of the area around PWN G54.1+0.3.
The first panel of Figure 2 provides a finding chart for the objects in Figure 1.
A large elliptical SNR-like shell (centered at $l=53.9^\circ$, $b=0.2^\circ$) with angular size 
of 30$^{\prime}\times$26$^{\prime}$, with major axis tilted at about 45$^\circ$ clockwise
with respect to the $l$ direction, is clearly seen. Several
 bright small compact sources are situated along the low-latitude part of the shell. A few
 filamentary structures are seen in the high-latitude part of the shell. The brightest 
extended source in the image is a known HII region G54.09-0.06 (Lockman 1989), and is 
just outside the low-latitude part of the shell. Another known HII region G53.64+0.24 is 
detected within the large radio shell.
  
Figure 1 (right) shows a new high resolution image at 8$\mu$m from the Galactic Legacy Infrared
 Midplane Survey Extraordinaire (GLIMPSE) Legacy Project (Benjamin et al. 2003), with the 1420
 MHz radio contours overlaid. % and a mid-IR image at 21 $\mu$m from the Midcourse Space Experiment (Price et al. 2001).
The image, centred on the radio shell G53.9+0.2, shows that bright thermal emission (an IR 
shell including the HII regions G53.64+0.24 and G54.09-0.06) just surrounds the large-scale
 1420 MHz shell (down to a limit of 0.4 mJy (5 $\sigma$) at 8$\mu$m), and that little thermal 
emission comes from the radio shell or from its interior. 

\section{HI Absorption and CO Emission}
We have searched the VGPS radial velocity range from -110 to 160 km/s for features in the HI
 emission which might be related to the morphologies of the large SNR-like shell or PWN
 G54.1+0.3. We have found no convincing HI emissions coincident with either object.

HI emission and absorption spectra were constructed for the extended sources PWN G54.1+0.3 
and HII region G54.09-0.06, and for the compact sources G53.83-0.06 and G54.1+0.1. 
  The filamentary emission of the radio shell is too faint to yield an HI absorption spectrum. 
The source regions were taken as rectangular areas covering the region of bright 1420
MHz continuum emission, and the background regions were larger rectangular regions (with
the source rectangle excluded) (see Tian et al. 2007 and Leahy \& Tian 2008 for a 
detailed description of the methods for HI spectrum analysis).
The resulting spectra are given in Fig. 2.
Also given in Figure 2 is the distance vs. radial velocity relation in the direction
$l$=54.1$^\circ$ for our adopted rotation curve (see the section on Kinematic Distances below).
For G54.1+0.3,  absorption features are seen all the way up to the tangent velocity, at 
$\sim$62 km/s, but no significant features at negative velocities are seen. 
We conclude that G54.1+0.3 lies beyond the tangent point but within the Solar circle in that direction.
G54.09-0.06 has a similar absorption spectrum to that of G54.1+0.3, i.e  
absorption features between sun and the tangent 
point and no absorption at negative velocities. Thus it is located at the far side of its 
recombination line at velocity ($\sim$42 km/s, Lockman 1989). Both G53.83-0.06 and G54.1+0.1
 are outside the Solar circle, since prominent absorption features appear at negative 
velocities, so they have nothing with PWN G54.1+0.3 or the large radio and IR shells.   

$^{13}$CO-line data was extracted from the survey of Jackson et al.(2006). 
The $^{13}$CO images show that a several clumps of molecular gas 
with radial velocity of $\sim$37 km/s are located just outside the large SNR-like shell (Fig.3a). 

We also find a molecular cloud with radial velocity of $\sim$53 
km/s is morphologically associated with PWN G54.1+0.3 (Fig. 3b) and thus should be at the same
distance os G54.1+0.3.
The $^{13}$CO spectrum in the direction of PWN G54.1+0.3 shows high-brightness emission with 
a width of $\sim$3 km/s centered at a radial velocity of 53 km/s. This may be explained as 
evidence of interaction between the PWN shock and this CO cloud. 
>From the CO emission spectrum in this direction, 
several CO clouds have HI absorption identified at the same velocity (see third panel of Fig. 2). 
This means these clouds are in front of G54.1+0.3 (except in the case of the 53 km/s
CO cloud which is located adjacent to G54.1+0.3, and only needs to be partially in front). 

\section{Discussion}  
\subsection{Kinematic Distances}

The flat Galactic circular rotation curve model is widely used in order to estimate kinematic 
distances to Galactic objects. Taking the commonly-used parameter of V$_{0}$ as 220 km/s 
(the circular velocity of the Sun), one finds a tangent point velocity, $V_T$ of 42 km/s in the 
direction of $l$=54.1$^{0}$. 
This is not consistent with the observed value of $V_T\simeq$62 km/s
 obtained from the HI emission spectra in Fig. 2. To obtain $V_T=$62 km/s, 
one finds V$_{R}\sim$240 km/s. 
Alternately, the high $V_T$ could be due to peculiar motions, such as spiral arm velocity 
perturbations, near the tangent point in the direction of $l$=54.1$^{0}$. 
A better knowledge of the rotation curve and 
non-circular motions is therefore needed in order to determine better distances to the 
Galactic objects. However, this is beyond the scope of this paper. 
For simplicity, hereafter, 
we calculate kinematic distance to an object by using a rotation curve model with 
 R$_{0}$= 7.6 kpc (the improved distance to the Galactic 
center from the Sun, Eisenhauer et al. 2005), V$_{0}$=220 km/s and  V$_{R}$ linearly 
increasing to 240 km/s as R decreases to its value at the tangent point ($R_T=6.16$kpc).
We note that Reid et al.(2007) has given an updated value for $\theta_0/R_0$ which results
in V$_{0}$=224 km/s and a similar value of V$_{R}$(243 km/s) at the tangent point.
Levine et al.(2008) also find a high value of rotation velocity ($\sim$236 km/s) 
at longitudes hear 53 degrees.
We estimate an uncertainty in distance
determination caused by streaming and random gas motions at $\sim$7 km/s (Shaver et al. 1982).% and non-circular motion ($\sim$ 5 km/s).

The most distant reliable absorption feature seen against G54.1+0.3 is at $\sim$62 km/s, which 
is at the tangent point, thus giving a lower limit distance of $\simeq$4.5 kpc. 
No HI absorption feature at negative velocities means that G54.1+0.3 is within the 
Solar circle, i.e. an upper limit distance of $\simeq$9.0 kpc. The morphological association
 between G54.1+0.3 and the CO cloud at $\sim$53 km/s suggests that G54.1+0.3 is at the 
far kinematic distance for the velocity 53$\pm$12 km/s, i.e. at a distance of 
6.2$^{+1.0}_{-0.6}$ kpc.  

Two extended HII regions G54.09-0.06 and G53.64+0.24 in the IR shell have been detected in
a previous recombination line survey at velocities of 42.1$\pm$0.8 km/s and 38$\pm$1.8 km/s, 
respectively. G54.09-0.06 is at the far side of its recombination line velocity due to the 
HI absorption feature at tangent point, i.e. 7.1$^{+0.7}_{-0.6}$ kpc. Absorption in the 
spectrum of G53.64+0.24 indicates that it is also at the far side of its recombination line
 velocity, i.e. at 7.5$^{+0.7}_{-0.6}$ kpc.  If the HII regions are actually part of the IR
 shell, then the IR Shell is likely at the same distance of $\sim$7.3 kpc.  

\subsection{Physical Relation Between the IR Shell, the Radio Shell and the PWN G54.1+0.3} 

Infrared images here show the large IR shell just surrounds the large 1420 MHz shell
and there is no IR emission within the IR shell interior. 
The coincidence of the IR shell morphology and
 the radio shell morphology indicates that the radio shell is inside the IR shell and
at same distance ($\sim$ 7.3 kpc). 

We next discuss whether the radio shell is thermal or non-thermal.
The radio shell had been described as a bright ring by previous 
low-resolution WSRT observations at 609 MHz (Velusamy et al. 1986). 
The WSRT data has no zero-spacing data, so that the fluxes are lower limits.
>From the WSRT image of the SNR shell and our 1420 MHz image, we find a lower limit to
spectral index of $\alpha>$-0.5 (with $S_{\nu}=S_0 \nu^{-\alpha}$), which is not useful
to distinguish thermal and non-thermal emission.
We also obtained the GB6 survey (Condon et al. 1991) 6cm image to determine 
1420-4850MHz spectral index.
The GB6 maps were made using a differencing technique, so they miss large scale flux. 
We calculate the spectral index of the bright filament at (l,b)=(53.75,0.30) and the
fainter filament at (l,b)=(54.10,0.35) and find in both cases upper limits of $\alpha <$ 1.7.
To check this result we calculate spectral index for the nearby supernova remnant G54.4-0.3,
which has a spectral index of 0.5 (Caswell, 1985) and find the same upper limit on $\alpha$. 
Thus the existing radio data does not give definitive evidence on whether the radio shell
has a thermal or non-thermal spectrum.

Velusamy et al. (1986) roughly classify the radio shell as an HII region G53.9+0.3 due to 
possible thermal IR emission from the shell. 
However our present images, which have higher spatial resolution, clearly show the radio 
and IR shells are physically separated, invalidating the previous conclusion. 
The radio shell might have a thermal radio spectrum and still show no IR emission 
if the dust in the inner side of the IR shell has mainly been destroyed, e.g., by intense UV 
radiation from young stars in the interior. Then the radio shell could be 
a stellar wind bubble produced by a group of young stars. 
PWN G54.1+0.3 could be associated with the same star cluster, but since the
the radial velocity of G54.1+0.3 is so different than than of the IR shell
(53 km/s vs. 40 km/s) we don't think that they are associated. 
Ignoring the velocity difference, and assuming G54.1+0.3
and the radio/IR shells are at the same distance, then we can explain the lack of an outer
shock for PWN G54.1+0.3 as being caused by the low density interior of the stellar wind bubble
inside the radio and IR shells. 

If the radio shell is non-thermal, then it is likely a SNR. 
G54.1+0.3 is $\sim$0.25$^\circ$ offset from the radio shell center, has a young age of $\sim$2000 
yrs (Lu et al 2002) and appears to be significantly nearer (6.1 kpc compared to 7.3 kpc). 
If we assume the pulsar in PWN G54.1+0.3 is born at the center
of the radio shell and at same distance as the radio (and IR) shell, the offset
corresponds to a distance of $\sim$32 pc, which gives an implausible high proper motion 
velocity of $\sim$16000 km/s for the pulsar. So, we conclude PWN G54.1+0.3 is highly unlikely 
to be the compact remnant of the same explosion that caused the SNR-like shell. 
Considering that the radio shell may be a new SNR G53.9+0.2, it has a distance of 
$\simeq$7.3 kpc and mean radius $\simeq$30 pc. The Sedov age is then 
80,000$(n_0/\epsilon_0)^{1/2}$yr, with $n_0$ the ambient density and $\epsilon_0$ the explosion
energy in units of 0.75$\times 10^{51}$ erg. 
For such a large SNR to still be in the Sedov phase requires a low ambient density or large 
explosion energy: $\epsilon_0 /n_0 > 4.9$ (e.g. using the limit that $t<t^{(d)}$, the time 
that the dynamics are affected by cooling, Cox, 1972). If the SNR is in the shell phase,
one can take the current radius of 30 pc as the cooling radius, to estimate a lower limit to
the age as the time when it enters the shell phase, $t^{(c)}=62000 n_0^{-37/85}$ yr. In
either case it is clear that G53.9+0.2 is an old SNR, and is not expected to be detected
in X-rays. We have verified there is no soft X-ray emission detected in the ROSAT 
All-Sky-Survey.

The distance we find for PWN G54.1+0.3 is 6.2$^{+1.0}_{-0.6}$ kpc, consistent but better
constrained that previous distance estimates. Camillo et al. (2002) discuss the energetics 
of the Crab-like PSR J1930+1852 in G54.1+0.3 using a distance of 5 kpc. 
The larger distance here implies
that the 2-10 keV X-ray luminosity is larger by a factor of 1.54, thus significantly reducing
the difference in luminosity with PWN/PSR J1124-5916, which has nearly identical spin
parameters to PSR J1930+1852.

\section{Conclusion}
Here we have searched for a SNR shell counterpart to the PWN G54.1+0.3 using radio continuum,
HI and CO line emission, and infrared observation. In addition we have used HI absorption and
CO emission data to constrain distances to PWN G54.1+0.3 and other nearby sources. We determine
a distance of $\simeq$6.2 kpc to G54.1+0.3 using an associated CO cloud, consistent with the HI
absorption spectrum which limits G54.1+0.3 to be between the tangent point and the far side 
of the solar circle. In addition, we find a SNR-like shell in 1420 MHz continuum 
emission, G53.9+0.2, which is offset from G54.1+0.3. 
The radio shell has no thermal IR emission from
its shell but is surrounded by an IR shell with an embedded HII region and associated CO 
clumps, both of which are located at 7.3 kpc distance. The radio shell 
may be thermal and lack IR emission due to dust destruction, or it may be nonthermal and
part of a newly found old SNR. In either case it is located at a different distance than 
PWN G54.1+0.3. 
 
\begin{acknowledgements}
DAL and TWW acknowledge support from the Natural Sciences and Engineering Research Council of Canada. 
TWW thanks the Natural Science Foundation of China for support.  
This publication makes use 
of molecular line data from the Boston University-FCRAO Galactic Ring Survey (GRS). The NRAO 
is a facility of the National Science Foundation operated under cooperative agreement by 
Associated Universities, Inc. 

\end{acknowledgements}

\begin{figure}
\vspace{70mm}
\begin{picture}(80,80)
\put(-20,250){\includegraphics{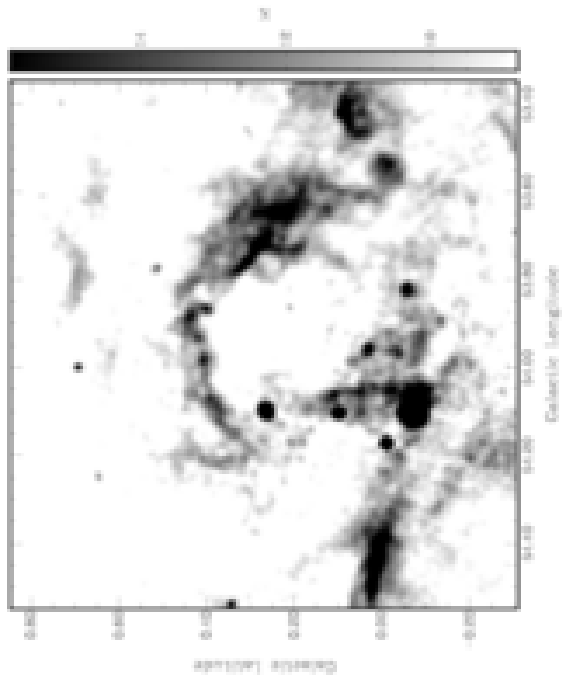}}
\put(230,280){\includegraphics{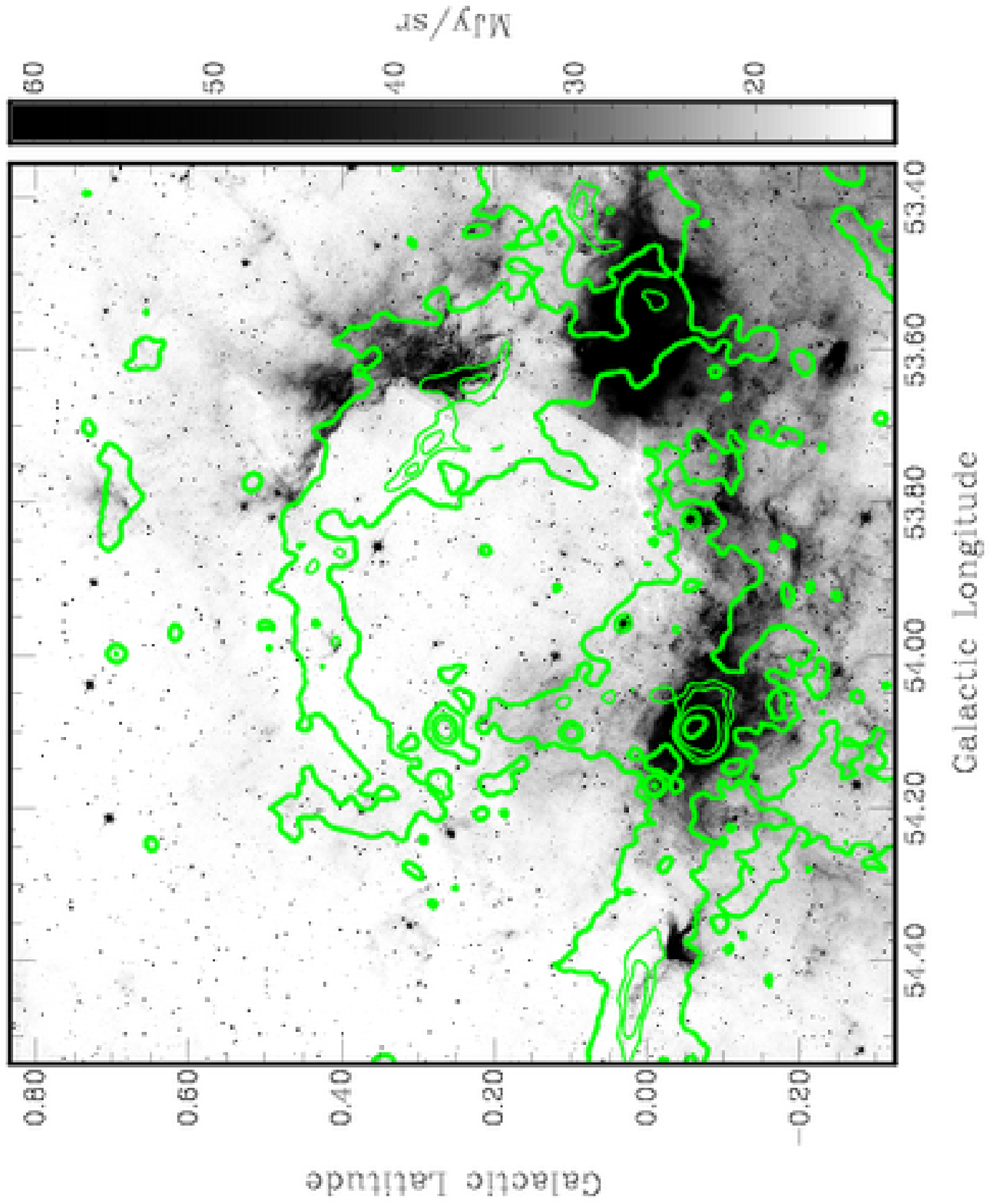}}
\end{picture}
\caption[xx]{Left: 1420 MHz continuum image centered at $l=53.95^\circ$ $b=0.25^\circ$ with a 
size of $1.2^{\circ}\times1.2^{\circ}$. The positions of PWN G54.1+0.3 and other sources are
shown in the first panel of Fig. 2. 
Right: 8 $\mu$m mid-infrared image of the same region from the GLIMPSE Legacy survey with resolution of $\le$ 2 arcsec and sensitivity of 0.4 mJy, and with contours (10, 15, 18, 25, 50)K of the 1420 MHz continuum emission.}
\end{figure}

\clearpage

\begin{figure}
\vspace{180mm}
\begin{picture}(80,80)
\put(-15,470){\includegraphics{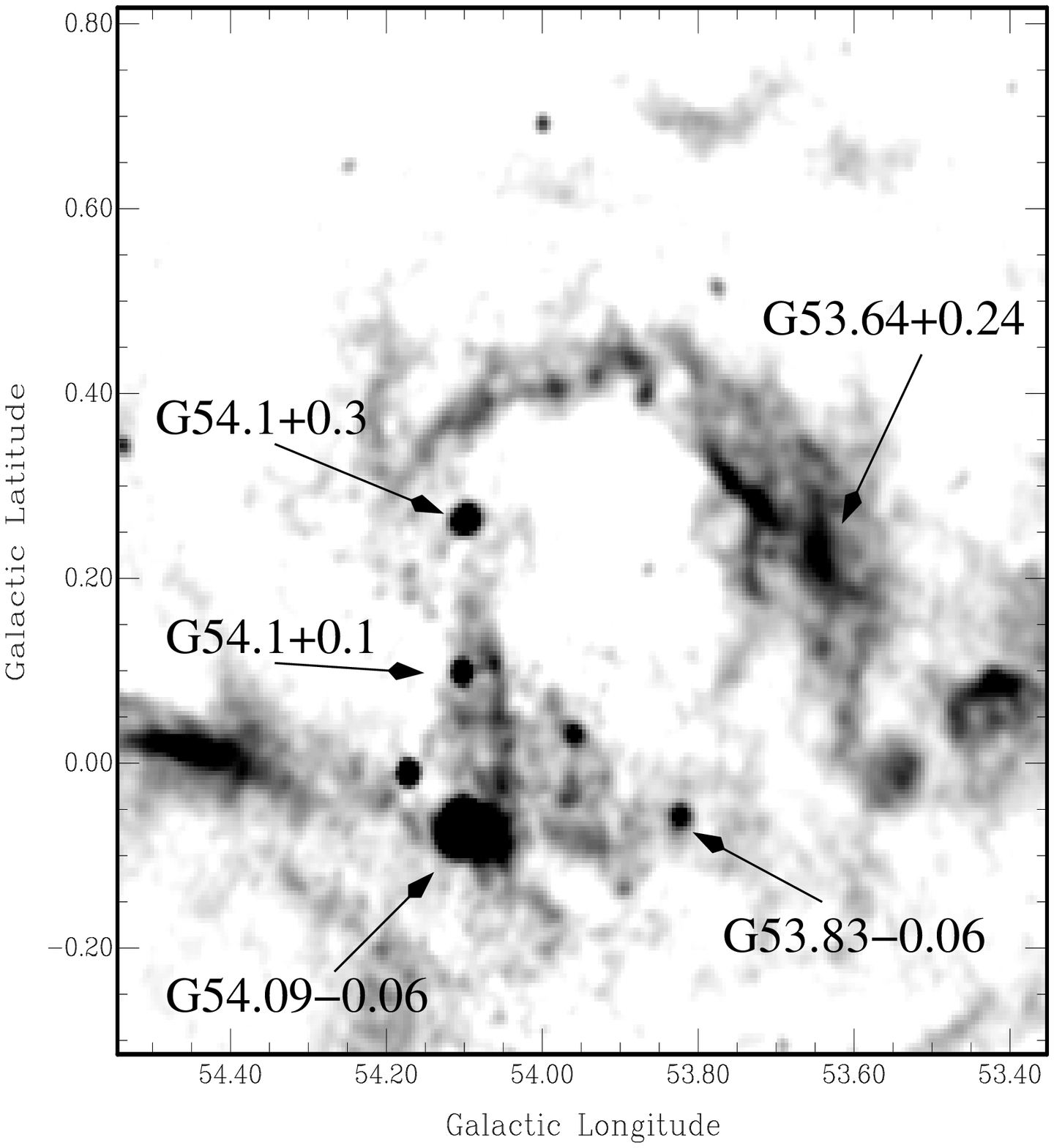}}
\put(225,455){\includegraphics{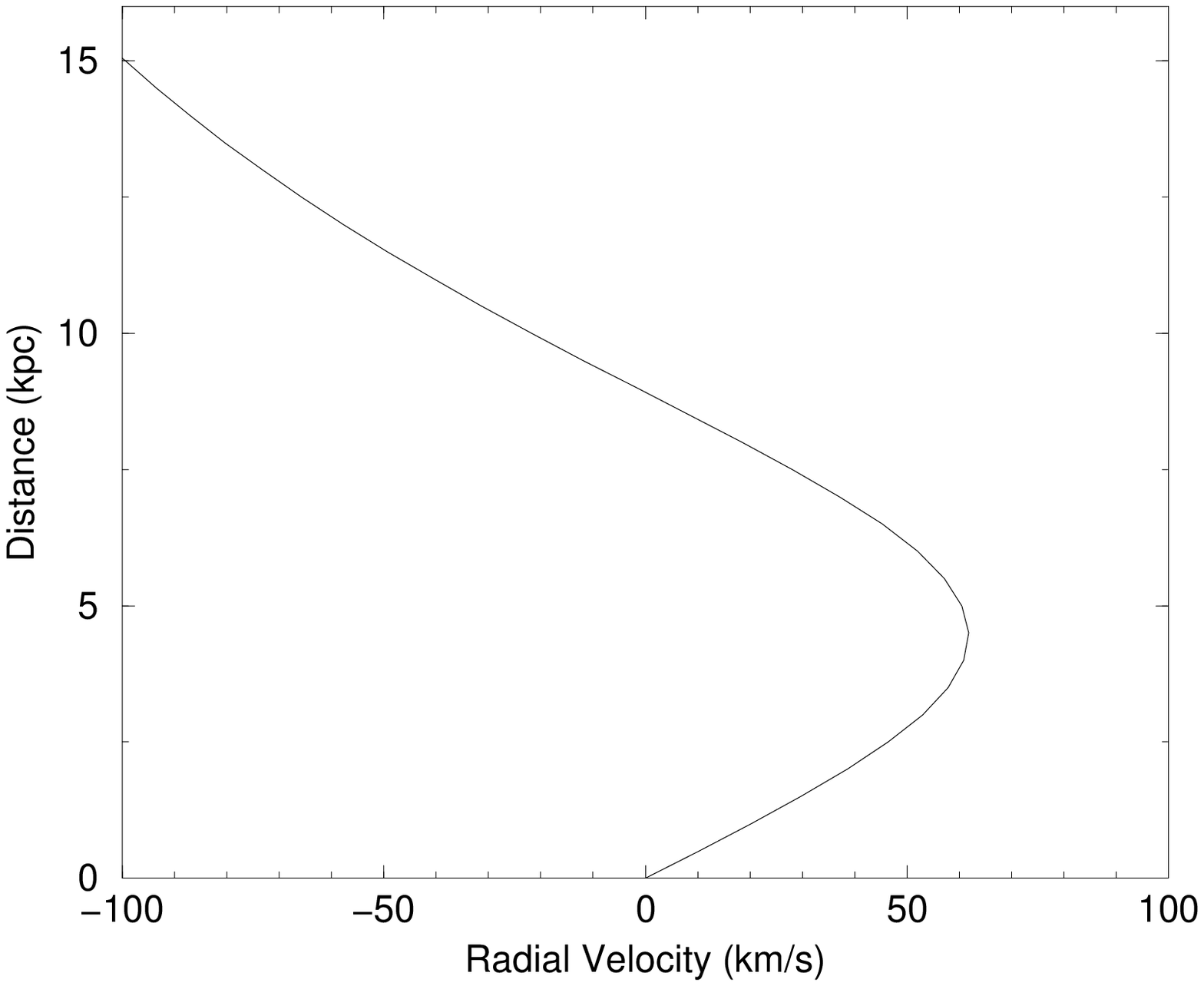}}
\put(-40,260){\includegraphics{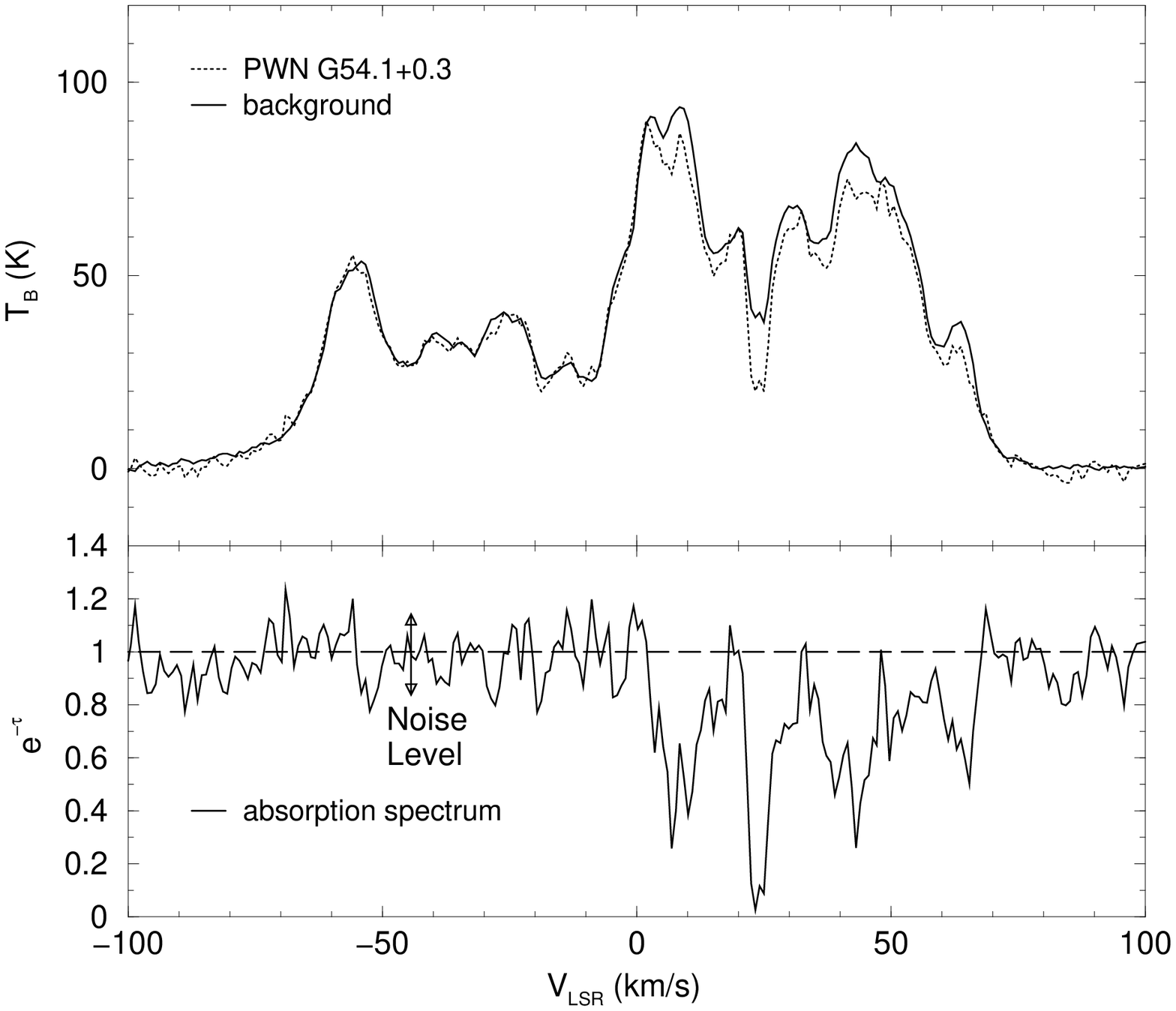}}
\put(230,260){\includegraphics{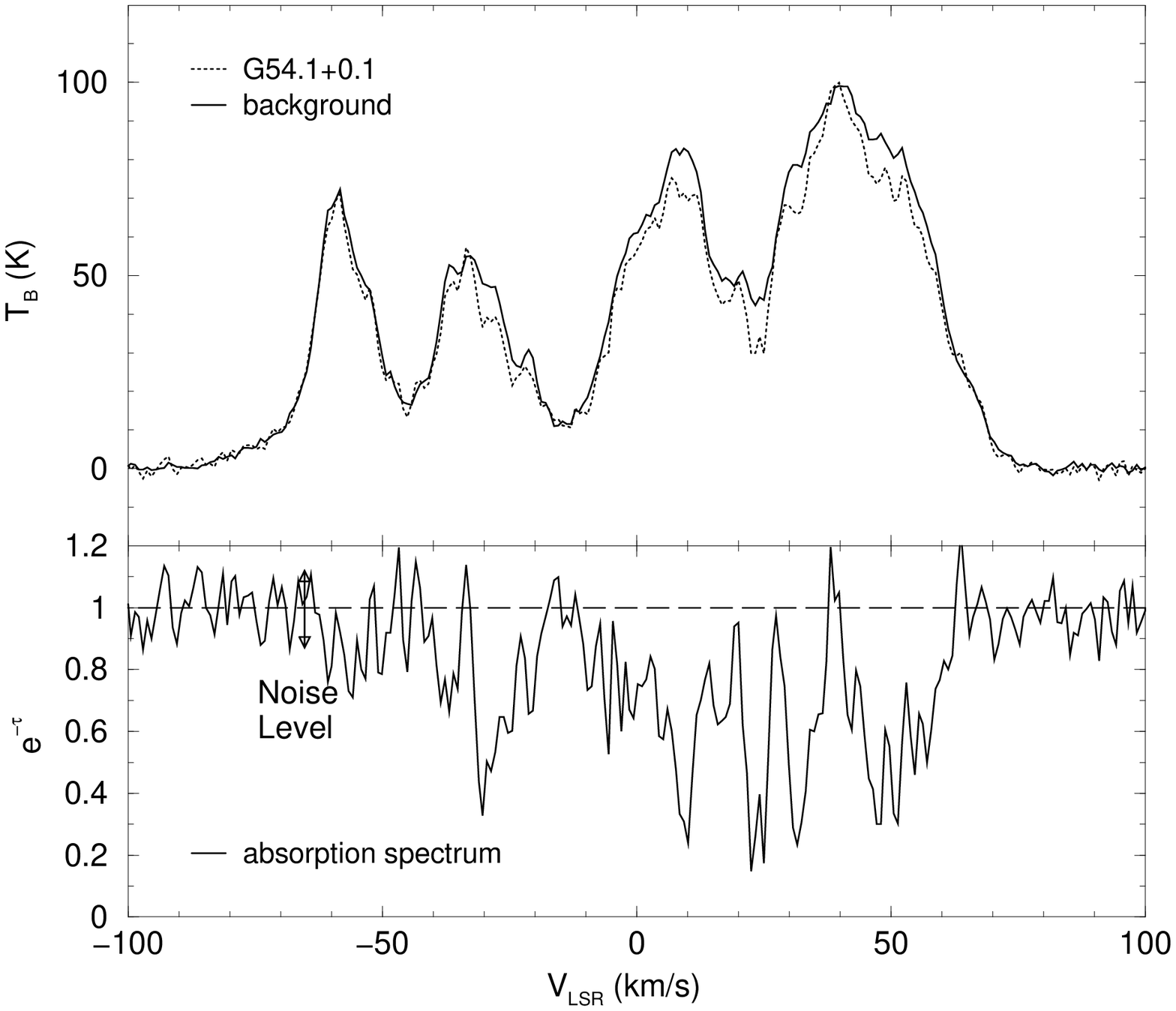}}
\put(-40,50){\includegraphics{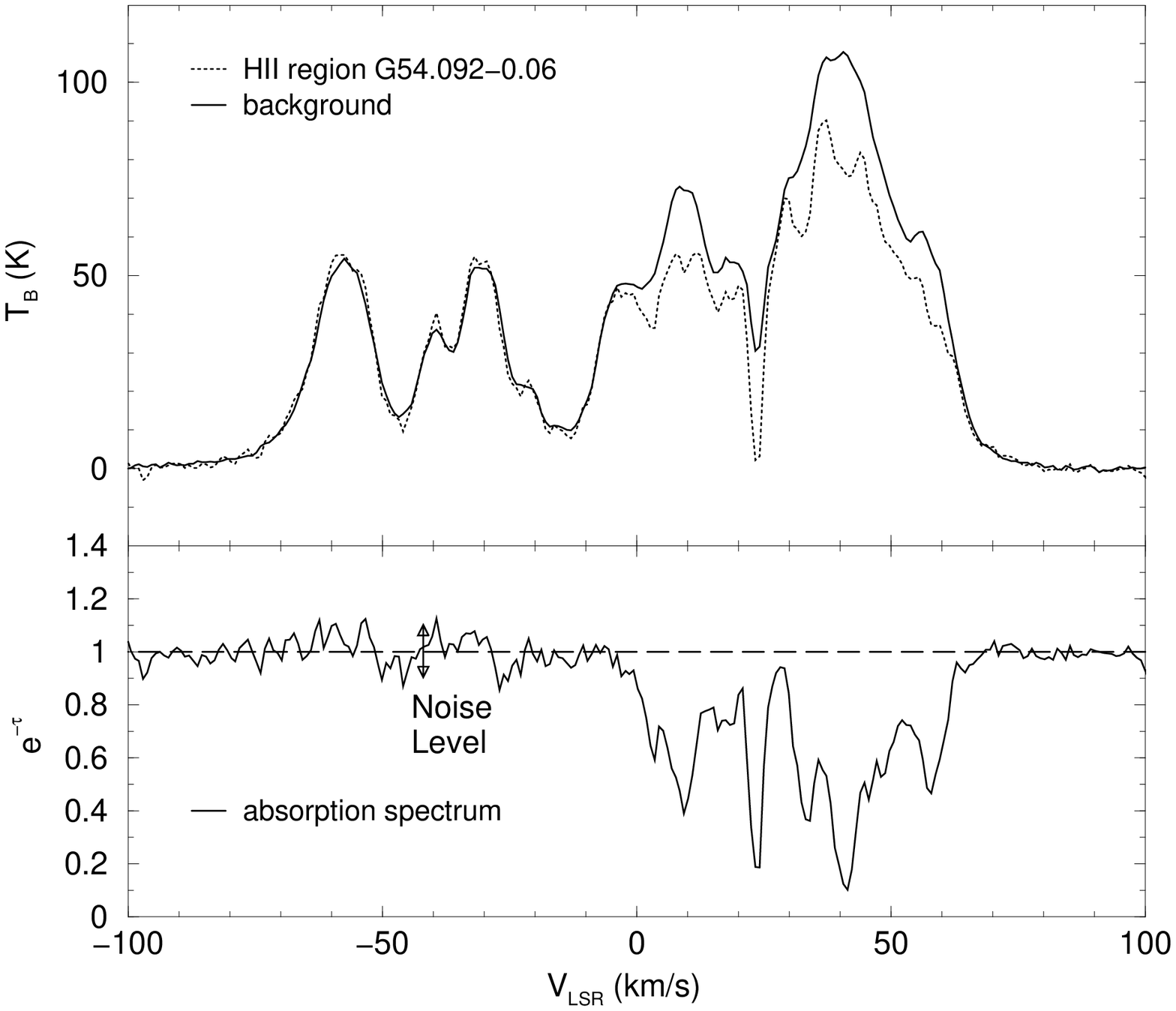}} \put(230,50){\includegraphics{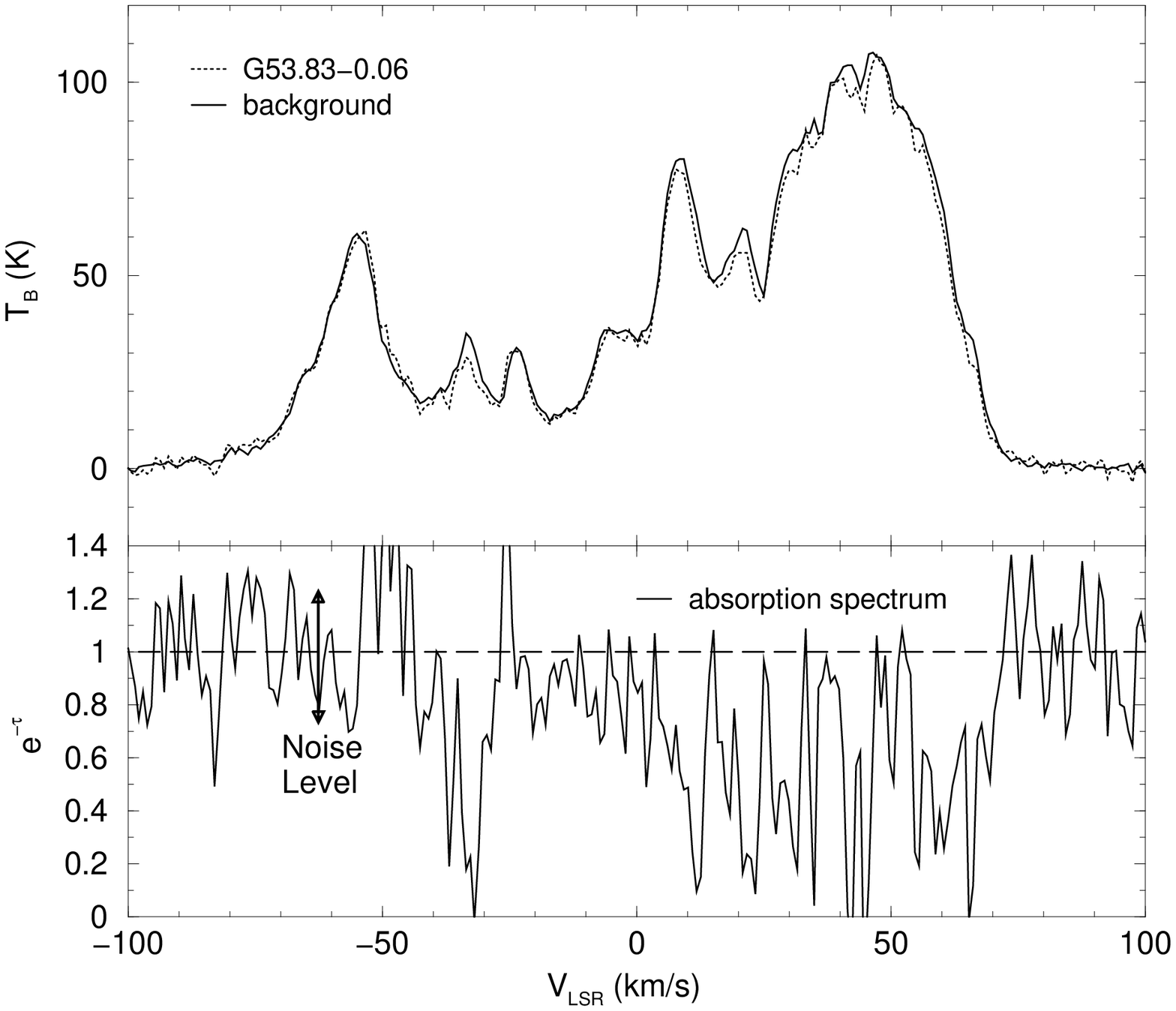}}
\end{picture}
\caption{The top left gives a finding chart for objects in Fig. 1. The top right shows 
the relation between distance and radial velocity for the assumed rotation curve.
The HI emission and absorption spectra of four bright sources are given in the
second and third rows.}
\end{figure}

\begin{figure}
\vspace{70mm}
\begin{picture}(80,80) 
\put(-20,50){\includegraphics{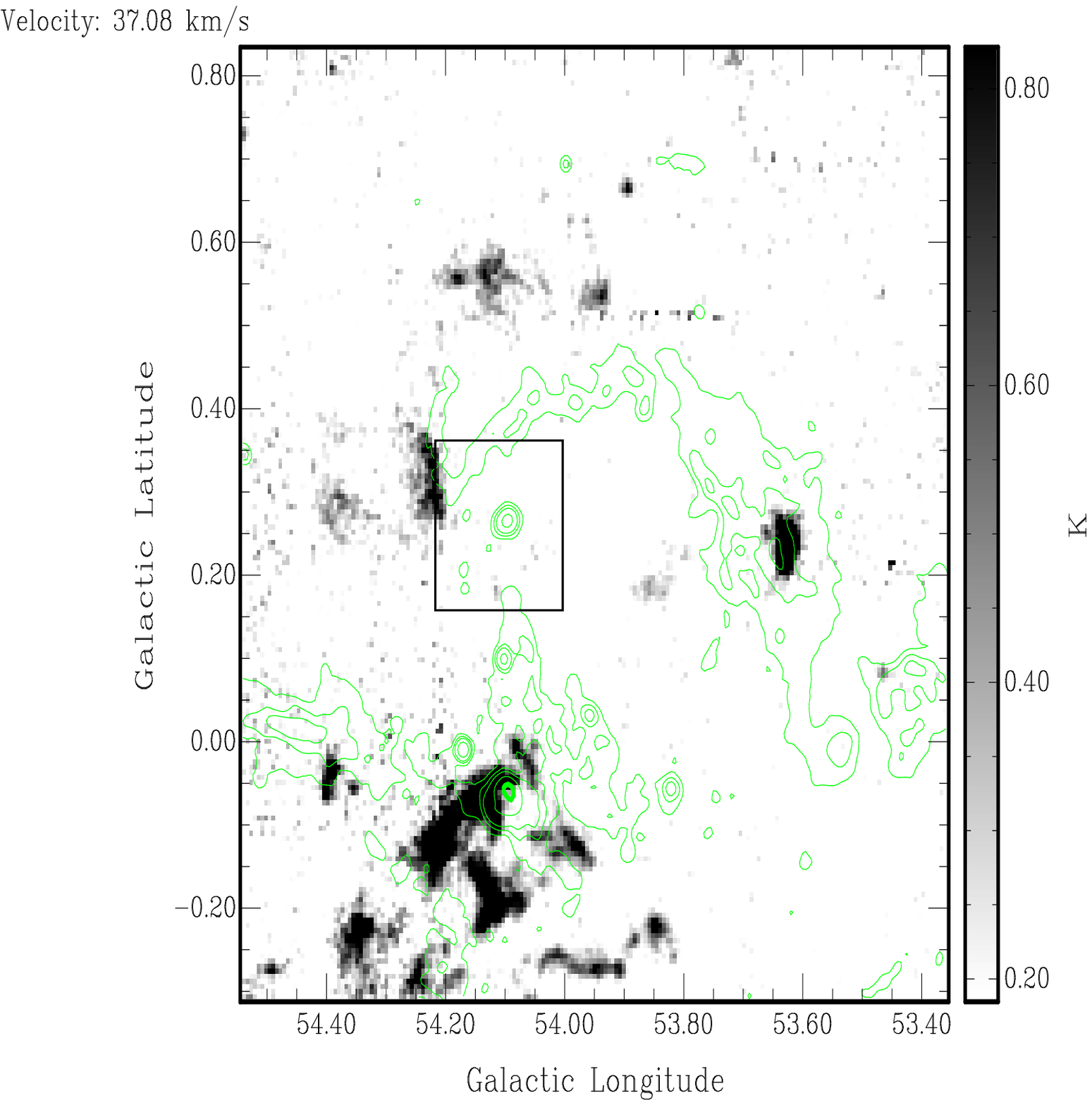}} 
\put(230,270){\includegraphics{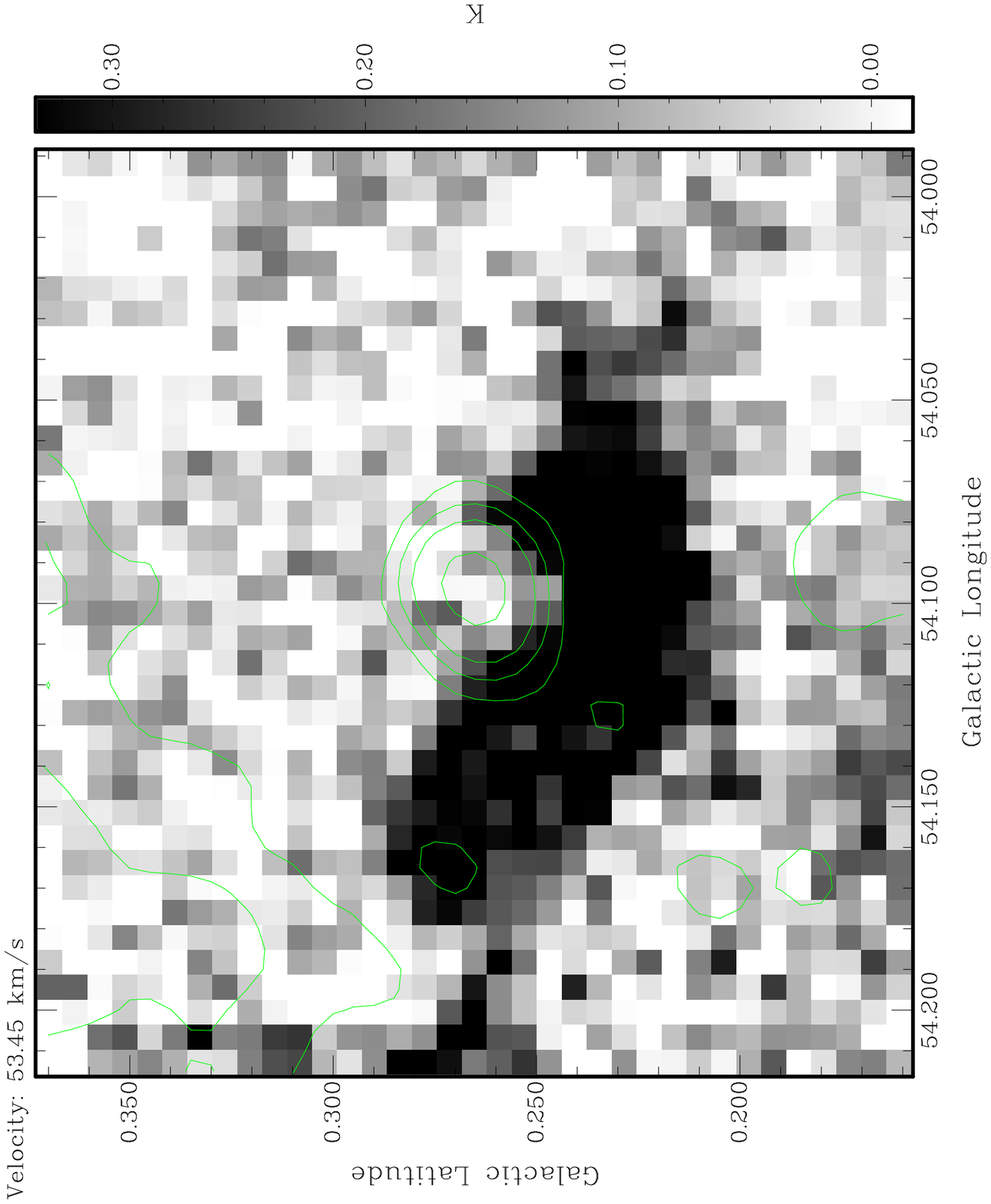}} 
\end{picture} 
\caption{The maps of CO emission at 37 km/s covering 1.2$^\circ\times 1.2^{\circ}$(left) 
and at 53 km/s covering 0.2$^\circ \times 0.2^{\circ}$(right). The rectangle in the left panel 
shows the area around PWN G54.1+0.3 covered by the right panel. These maps have superimposed 
contours (11 14 18 30 60 K) of the 1420 MHz continuum emission.} 
\end{figure}

\clearpage


\begin{thebibliography}{}
\bibitem[2003]{Benet03}Benjamin, R.A., Churchwell, E., Babler, B.L. et al. 2003, PASP, 115, 953
\bibitem[Camilo et al.(2002)]{2002ApJ...574L..71C} Camilo, F., Lorimer, 
D.R., Bhat, N.D.R., et al. 2002, ApJ, 574, L71 
\bibitem[Caswell 1985]{cas85}Caswell, J. 1985 AJ 90 1224
\bibitem[Condon et al.(1991)]{1991AJ....102.2041C} Condon, J.~J.,
Broderick, J.~J., \& Seielstad, G.~A.\ 1991, AJ, 102, 2041 
\bibitem[2002]{Coret02}Cordes, J.M. \& Lazio, T.J.W. 2002, astroph/0207156 
\bibitem[1972]{Cox72}Cox, D.P. 1972, ApJ, 178, 159 
\bibitem[2005]{Eiset05}Eisenhauer, F., Genzel, R., Alexander, T. et al. 2005, ApJ, 628, 246
\bibitem[1995]{Fraet95}Frail, D.~A., Kassim, 
N.E., Cornwell, T.J., \& Goss, W.M. 1995, ApJ, 454, L129 
\bibitem[2006]{Jaket06}Jackson, J.M., Rathborne, J.M., Shah, R.H. et al. 2006, ApJ Suppl., 163, 145 
\bibitem[2008]{Leaet07}Leahy, D.A., Tian, W.W. 2008, AJ, 135, 167
\bibitem[2008]{Lev08} Levine, E.~S., Heiles, C., \& Blitz, L.\ 2008, ApJ, 679, 1288 
\bibitem[1989]{Loket89}Lockman, F, J. 1989, ApJS, 71, 469
\bibitem[2002]{Luet02}Lu, F.J. Wang, Q.D., Aschenbach, B. et al. 2002, ApJ, 568, L49
\bibitem[Reid et al.(2007)]{2007IAUS..242..348R} Reid, M.~J., Brunthaler, A., Menten, K.~M., Ye, X., Xing-Wu, Z., \& Moscadelli, L.\ 2007, IAU Symposium, 242, 348
\bibitem[1982]{Shaet82}Shaver, P.A., Radhakrishnan, V.,
 Anantharamaiah, K.R., et al. 1982, A\&A, 106, 105 
\bibitem[2006]{Stiet06}Stil, J.M., Taylor, A.R., Dickey, J.M. et al. 2006, AJ, 132, 1158
\bibitem[2007]{Tiaet07}Tian, W.W., Leahy, D.A., Wang, Q.D. 2007, A\& A, 474, 541
\bibitem[2008]{Tiaet08}Tian, W.W., Leahy, D.A., Haverkorn, M. Jiang, B. 2008, ApJ, 679L, 85
\bibitem[1988]{Velet88}Velusamy, T.\& Becker, R.H. 1988, AJ, 95, 1162
\bibitem[1986]{Velet86}Velusamy, T. Goss, W.M. \& Arnal, E.M. 1986, A\& A, 7, 105
\bibitem[Weisberg et al.(2008)]{2008ApJ...674..286W} Weisberg, J.M., 
Stanimirovi{\'c}, S., Xilouris, K., et al. 2008, \apj, 674, 286 

\end{thebibliography}
\end{document}